\documentclass[%
 reprint,
 amsmath,amssymb,
 aps,
]{revtex4-1}

\usepackage{graphicx}
\usepackage{dcolumn}
\usepackage{bm}

\begin{document}

\preprint{APS/123-QED}

\title{Experimental Test of the Relation between Coherence and Path Information}

\author{Jun Gao,$^{1,2}$ Zhi-Qiang Jiao,$^{1,2}$ Cheng-Qiu Hu,$^{1,2}$ Lu-Feng Qiao,$^{1,2}$ Ruo-Jing Ren,$^{1,2}$ Hao Tang,$^{1,2}$
Zhi-Hao Ma,$^{3}$ Shao-Ming Fei,$^{4}$ Vlatko Vedral,$^{5,6}$ }
\author{Xian-Min Jin$^{1,2,}$}
\email{xianmin.jin@sjtu.edu.cn}

\affiliation{$^1$State Key Laboratory of Advanced Optical Communication Systems and Networks, School of Physics and Astronomy, Shanghai Jiao Tong University, Shanghai 200240, China}
\affiliation{$^2$Synergetic Innovation Centre of Quantum Information and Quantum Physics, University of Science and Technology of China, Hefei, Anhui 230026, China}
\affiliation{$^3$School of Mathematical Sciences, Shanghai Jiao Tong University, Shanghai 200240, China}
\affiliation{$^4$School of Mathematical Sciences, Capital Normal University, Beijing 100048, China}
\affiliation{$^5$Centre for Quantum Technologies, National University of Singapore, 3 Science Drive 2, 117543, Singapore}
\affiliation{$^6$Clarendon Laboratory, University of Oxford, Parks Road, Oxford OX1 3PU, UK}

\date{\today}

\pacs{Valid PACS appear here}
\maketitle
\textbf{Quantum coherence stemming from the superposition behaviour of a particle beyond the classical realm, serves as one of the most fundamental features in quantum mechanics. The wave-particle duality phenomenon, which shares the same origin, has a strong relationship with quantum coherence. Recently, an elegant relation between quantum coherence and path information has been theoretically derived. Here, we experimentally test such new duality by $l_1$-norm measure and the minimum-error state discrimination. We prepare three classes of two-photon states encoded in polarisation degree of freedom, with one photon serving as the target and the other photon as the detector. We observe that wave-particle-like complementarity and Bagan's equality, defined by the duality relation between coherence and path information, is well satisfied. Our results may shed new light on the original nature of wave-particle duality and on the applications of quantum coherence as a fundamental resource in quantum technologies.}\\

\subsection*{Introduction}
\noindent Coherence was recognized early as a superposition of optical fields in the theory of electromagnetic waves. Together with the energy quantization, a quantum version of coherence has become one of the most fundamental features that can mark the departure of quantum mechanics from the classical realm \cite{1,2,3,4,Barbieri2009}. Carrying out general quantum operations remotely under local operations and classical communication requires quantum states that contain consumable resources. Quantum entanglement is found to have strong connection with coherence \cite{ChitambarSRB16,StreltsovCRB16, Killoran16} and may even originate from it \cite{StreSDB15}.The development of quantum technologies demands a reassessment of fundamental resources such as quantum coherence, including relations with other quantum physical phenomena and rigorous quantitative description \cite{Baumgratz,napoli,jiajunma,yrzhang,shumingcheng,uttam,rana,Manabendra}.

Among these quantum physical phenomena, wave-particle duality has been a unified picture to fully describe the behavior of quantum-scale objects. Quantitative characterizations of such wave-particle duality relations have been extensively investigated, aiming to set an upper bound on the sum of the wave behavior and the particle behavior for a given interferometer \cite{3,4}. Based on the fringe visibility $V$ of the interference pattern and the path distinguishability $D$, a wave-particle duality relation is given by $V^2 + D^2 \leq 1$, which means that full wave-behavior ($V = 1$) implies no particle behavior ($D = 0$) and vice-versa.

Recently, Bagan {\it et.al} \cite{EB} proposed another elegant relation to characterise the new wave-particle-like duality based on the coherence $C^{l_1}$ which quantifies the wave nature of a state and the path information which is given by the minimum-error state discrimination between the detector state and the target state. In this paper, we experimentally test such new duality relation by $l_1$-norm measure and the minimum-error state discrimination. By preparing three classes of two-photon states that have different upper bounds of quantum coherence, we are able to investigate the new wave-particle-like complementarity in different regions. In every class of state, we continuously tune the detector state and observe clear duality trade-off as well as the upper bound on the quantum coherence of the target photon. The Bagan's equality defined by the duality relation between coherence and path information can be well satisfied.

\begin{figure*}
 \centering
 \includegraphics[width=2.0\columnwidth]{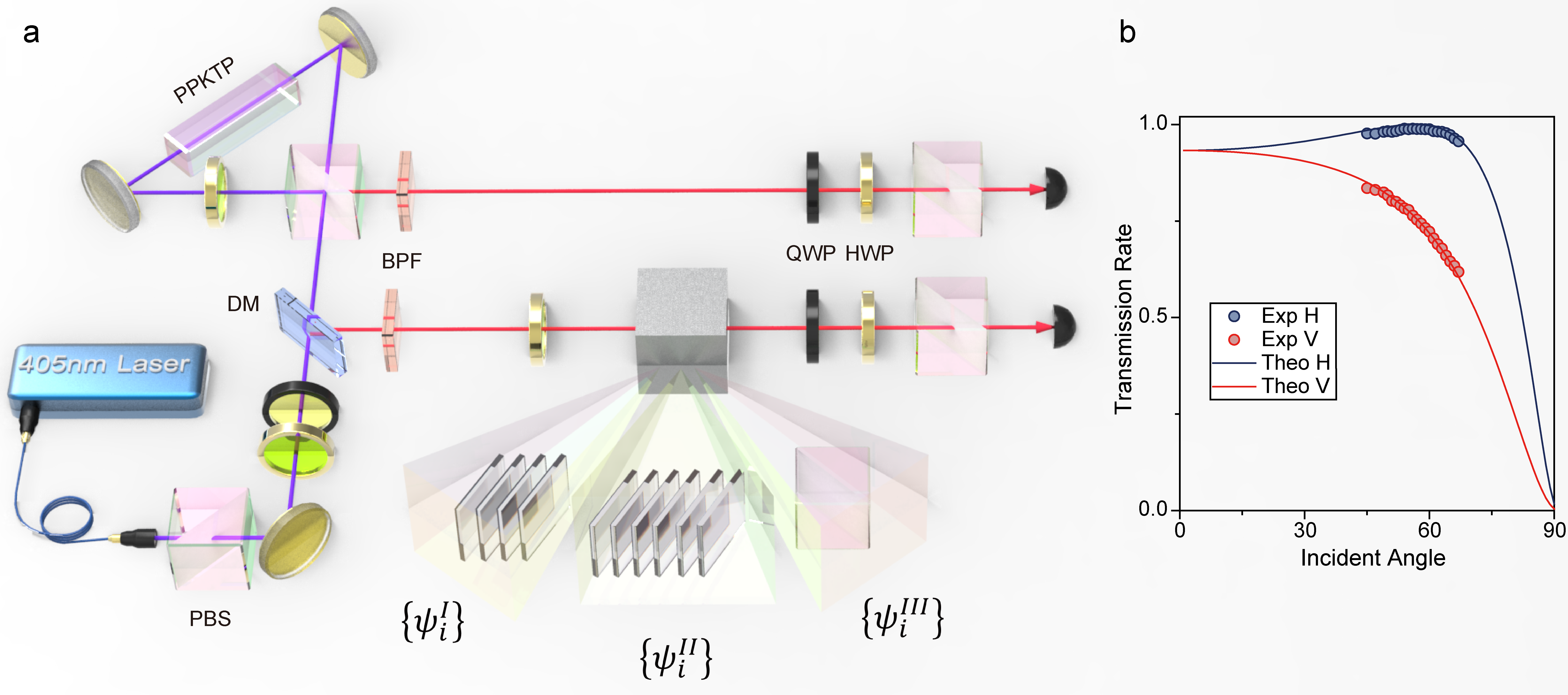}\\
 \caption{\textbf{Schematic view of experiment.} \textbf{a.} Experimental setup. Pairs of polarisation entangled photons are generated via spontaneous parametric downconversion in a $25$mm long periodically-poled Potassium Titanyl Phosphate (PPKTP) crystal pumped by a $405$nm UV laser with Sagnac loop scheme. After locally rotating the polarisation of one path, the photons are guided into the state generation section. By either inserting different pieces of Brewster window or place a polarisation beam splitter (PBS), three different classes of states $\{\psi^{I}_i\}$, $\{\psi^{II}_i\}$ and $\{\psi^{III}_i\}$ are generated. The states are then analysed by the polarisation analysis measurement setup composed of a half-wave plate (HWP), a quarter-wave plate (QWP) and a PBS. \textbf{b.} The Brewster window characterization. Transmission rate of fused silica is characterised with different incidence angles and polarisations. The results suggest $60^\circ$ as a reasonable incidence angle.}
 \label{FIG. 1.}
\end{figure*}

\subsection*{Results}
\textbf{Theoretical description of the duality relation.} Consider a particle (target state) entering an N-port interferometer via a generalized beam splitter. Once the particle interacts with the detector (detector state), the state of the entire system is described by
\begin{equation}\label{pd}
\arrowvert{\psi}\rangle=\sum^{N}_{i=1}\sqrt{p_i}\arrowvert{i}\rangle\arrowvert{\eta_i}\rangle,
\end{equation}
which is the superposition of the initial target particle state $\arrowvert{i}\rangle$ and the detector state $\arrowvert{\eta_i}\rangle$. The coherence of the target state is given by
$C=C^{l_1}(\rho)/N$, where $\rho=Tr_{det}(\arrowvert{\psi}\rangle\langle{\psi}\arrowvert)
=\sum^{N}_{i,j=1}\sqrt{p_ip_j}\langle{\eta_j}\arrowvert{\eta_i}
\rangle\arrowvert{i}\rangle\langle{j}\arrowvert$.
$C^{l_1}(\rho)=\sum_{i\neq j} |\rho_{ij}|$, where $|\rho_{ij}|$ is the absolute value of entry $\rho_{ij}$ of the density matrix $\rho$ under the reference basis\cite{Baumgratz}. Here, $C$ quantifies the wave nature of the target particle.

\begin{figure*}
 \centering
 \includegraphics[width=2.0\columnwidth]{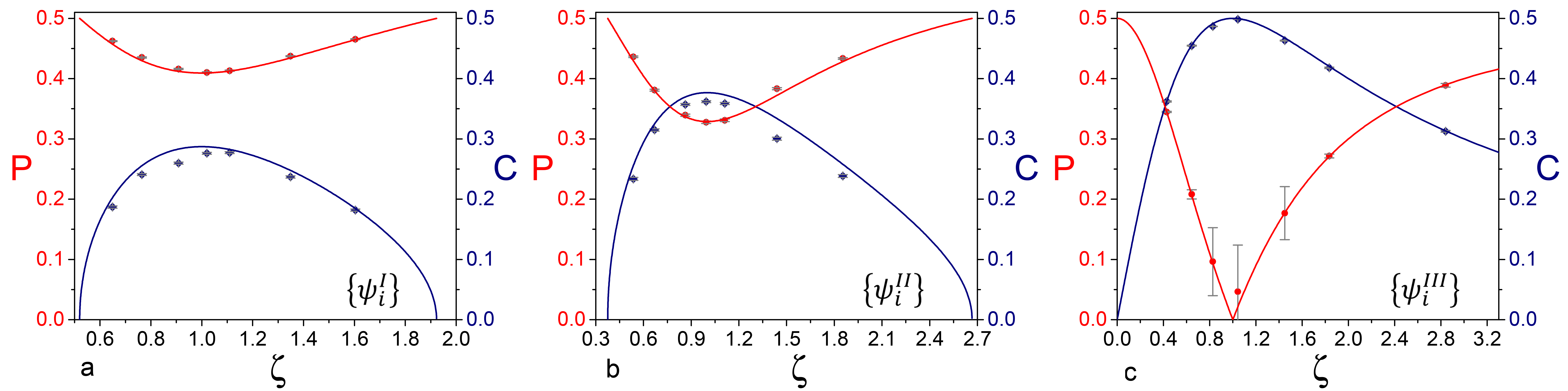}\\
 \caption{\textbf{Experimental observation of the new wave-particle-like duality with three different classes of states.} From \textbf{a} to \textbf{c}, in each class of states, as we vary the parameter $\zeta$, we experimentally observe this new relation trade-off, namely, a continuous transition between quantum coherence and path information. The blue diamonds represent the coherence properties while the red dots give the path information. As we vary the states from $\{\psi^{I}_i\}$ to $\{\psi^{III}_i\}$, the degree of complementarity also increases. The solid curves are the theoretical results based on the three different classes of states. Error bars are calculated with Monte Carlo simulation with the Poissonian statistics of the detection process taken into account.}
 \label{FIG. 2.}
\end{figure*}

The detector states are introduced to quantify the path information of the target particle by tracing out the target particle state,
$\rho_{det}=Tr_{tar}(\arrowvert{\psi}\rangle\langle{\psi}\arrowvert)
=\sum^{N}_{i=1}p_i\rho_i$, where $\rho_i=\arrowvert{\eta_i}\rangle\langle{\eta_j}\arrowvert$.
If the detector states $\rho_i$ are orthogonal, no coherence property of the target particle will be obtained.
To discriminate among the detector states $\arrowvert{\eta_i}\rangle$, one employs the minimum-error strategy by using an $N$-element positive operator valued measure (POVM)
with elements $\{\Pi_i\}$. Then the average probability of successfully identifying the state is $P_s=\sum_{i=1}^N p_i \langle{\eta_j}\arrowvert\Pi_i\arrowvert{\eta_i}\rangle$. It is proved that $C$ and $P_s$ satisfy the following new relation raised by Ref.\cite{EB},
\begin{equation}\label{thm}
(P_s-\frac{1}{N})^{2}+C^{2}\le(1-\frac{1}{N})^{2}.
\end{equation}
This upper bound represents a trade-off between the path information and the coherence of the target particle. In our experiment, we consider the case of $N=2$. It is worth mentioning that when $N=2$, the inequality becomes an equality which we denote as Bagan's equality. In this case, $P_s$ can be maximised either by an optimised POVM or calculated by an analytic solution\cite{Qiu}. It should be noticed that this new definition of path information is different from Ref.\cite{3}. Here, we use the success probability in minimum-error state discrimination, while Ref.\cite{3} uses the difference of two probabilities for taking either one of the two paths.

\textbf{Experimental test of the duality relation.}We experimentally generate the two-photon polarisation entangled state via type-II spontaneous parametric down-conversion\cite{PPKTP}. As shown in Fig.1a, a $405$nm UV laser is first coupled into a single mode fiber to acquire a high quality spatial beam profile. After adjusting the polarisation of the pump laser with a combination of a half-wave plate (HWP) and a quarter-wave plate (QWP), the pump laser enters the Sagnac interferometer and is focused on a periodically-poled Potassium Titanyl Phosphate (PPKTP) crystal. The clockwise and the anti-clockwise components then interfere at the central polarisation beam splitter (PBS) and generate a superposition state of these two components. The generated singlet state can be written as $\arrowvert{\psi^{-}}\rangle=\frac{1}{\sqrt{2}}(\arrowvert{H_1V_2}\rangle-\arrowvert{V_1H_2}\rangle)$. Two band pass filters (BPF) centred at the desired wavelength of the down-converted photons are employed to block the UV laser. With careful alignment of the Sagnac interferometer, we are able to observe the polarisation visibility as high as $97.7\%$ in $H/V$ basis and $98.3\%$ in $D/A$ basis. Then we further characterise the entanglement quality by conducting state tomography\cite{tomo}. The measured purity and concurrence of the entangled state is $0.963\pm0.0016$ and $0.964\pm0.0016$ respectively, which suggests that we have obtained a very pure state with high entanglement quality.

In order to observe the coherence properties of the target states, the detector states $\eta_i$ should be non-orthogonal. In our experiment, we first use a HWP to generate the following two-qubit state with four components,
\begin{equation}
\arrowvert{\psi}\rangle=\alpha\arrowvert{H_1H_2}\rangle+\beta\arrowvert{H_1V_2}\rangle+\gamma\arrowvert{V_1H_2}\rangle+\delta\arrowvert{V_1V_2}\rangle,
\end{equation}
where $\alpha$, $\beta$, $\gamma$ and $\delta$ are parameters based on the rotation in Hilbert Space. Note that these parameters are highly correlated and can be determined by the rotation angle of HWP and the normalization condition. Since $\arrowvert{\psi^{-}}\rangle$ state is invariant under arbitrary unitary rotation in Hilbert space, polarisation-dependent loss should be additionally introduced to generate non-orthogonal basis of the detector states. Here, we choose fused silica as Brewster window to introduce this polarisation-dependent loss into the above four components.

We experimentally characterise the transmission rate of fused silica with different incidence angles and the results agree well with the theoretical prediction (see Fig.1b). We choose $60^\circ$ as the incidence angle where the transmission rate is 99.7\% and 71.9\% for horizontal and vertical polarisation respectively (denoted as $\epsilon_h$ and $\epsilon_v$). After passing through this state preparation section the states can be described as
\begin{equation}
\begin{aligned}
\arrowvert{\psi}\rangle=&\arrowvert{H_t}\rangle(\alpha\epsilon^n_h\arrowvert{H_d}\rangle+\beta\epsilon^n_v\arrowvert{V_d}\rangle)+\\
&\arrowvert{V_t}\rangle(\gamma\epsilon^n_h\arrowvert{H_d}\rangle+\delta\epsilon^n_v\arrowvert{V_d}\rangle)
\end{aligned}
\end{equation}
where index $n$ represents the number of Brewster windows we insert, and subscripts $t$ and $d$ represent the target and detector qubits respectively. We can see that the detector qubit state $(\alpha\epsilon^n_h\arrowvert{H_d}\rangle+\beta\epsilon^n_v\arrowvert{V_d}\rangle)$ is not orthogonal to $(\gamma\epsilon^n_h\arrowvert{H_d}\rangle+\delta\epsilon^n_v\arrowvert{V_d}\rangle)$, and their inner product can be tuned with the rotation angle of HWP and the number of Brewster window. We are therefore able to test Bagan's theory in a wide range, a transition from the case of minimal coherence and maximal path information to the opposite case. 

\begin{figure}[!htb]
 \centering
 \includegraphics[width=1\columnwidth]{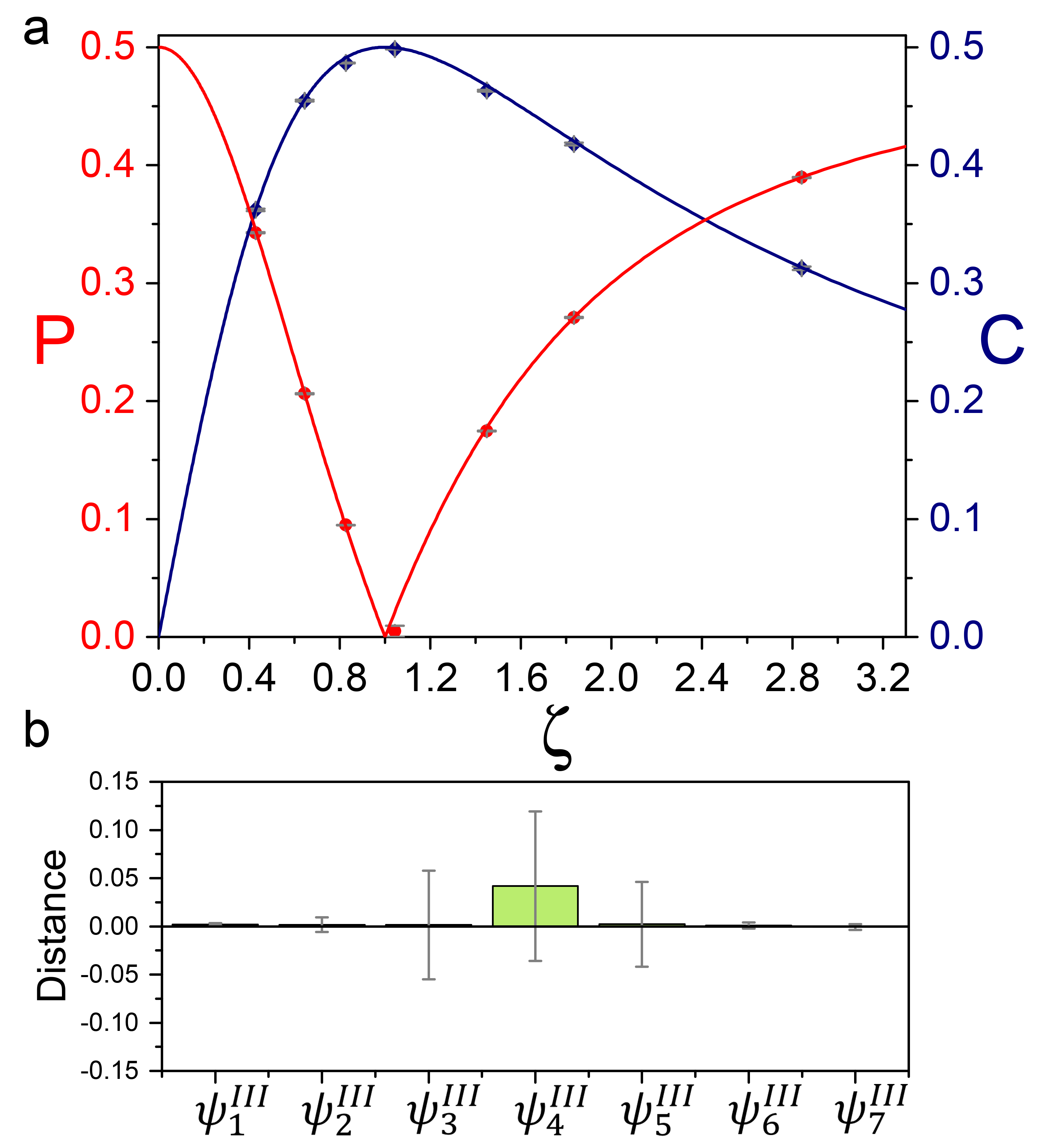}\\
 \caption{
  \textbf{Direct measured results of path information $P$ and the comparison result.} \textbf{a.} Coherence of $l_1$-norm measure and path information given by direct optimised positive operator valued measure (POVM) results of the third class state $\{\psi^{III}_i\}$. The solid curves are the theoretical results. \textbf{b.} A comparison of $P_s$ result between the analytic solutions and optimised POVM. Error bars in Fig.3 are directly calculated by using the propagation of error formulas.}
 \label{FIG. 3.}
\end{figure}

We then define a simplified parameter $\zeta$ as
\begin{equation}
\zeta=\sqrt{\frac{\alpha^{2}\epsilon_h^{n2}+\beta^{2}\epsilon_v^{n2}}{\gamma^{2}\epsilon_h^{n2}+\delta^{2}\epsilon_v^{n2}}}.
\end{equation}
From this expression we can see that parameter $\zeta$ is determined by the rotation angle of the HWP and the transmission rates for different polarisations. Once we have fixed the rotation angle and the number of Brewster window, the state we generated is also determined. To be more specific, $\epsilon^n_h$ and $\epsilon^n_v$ can set the upper bound on the quantum coherence of the states, and the rotation angle of the HWP (tune $\alpha$, $\beta$, $\gamma$ and $\delta$) can affect the duality trade off. As shown in Fig.1a, by inserting 4 and 6 pieces of Brewster window (affect $\epsilon^n_h$ and $\epsilon^n_v$), we are able to generate two different classes of states which we denote as $\{\psi^{I}_i\}$ and $\{\psi^{II}_i\}$ respectively. Here subscript ``i'' in the states corresponds to different HWP angles. Once $\epsilon^n_h$ and $\epsilon^n_v$ are fixed, then by tuning the angles of the HWP, we are able to observe the complementarity of this new duality relation. As we change the pieces of Brewster window from 4 to 6, the concurrence of the entangled state decreases from $0.795\pm0.0039$ to $0.650\pm0.0062$ while the purity of the states remains at a high level, $0.962\pm0.0035$ and $0.965\pm0.0041$ respectively. For each class of states, while we change the parameter $\zeta$ (by tuning different HWP angles), the entanglement and purity are not affected. This can be easily understood since local operations do not destroy the entanglement. For an asymptotic limit case, we use a PBS to replace the Brewster window to generate a separable state ($\{\psi^{III}_i\}$). With $\{\psi^{III}_i\}$, we can observe full complementarity between quantum coherence and path information. It should be noticed that the three classes of states are not the entire states of the system but part of states retrieved in a smaller subspace with postselection. Such solution has been proven to be exactly equivalent to the ideal one by Sandu Popescu, Lucien Hardy, and Marek Zukowski \cite{Sandu,Pan}. 

\begin{figure}[!htb]
 \centering
 \includegraphics[width=1\columnwidth]{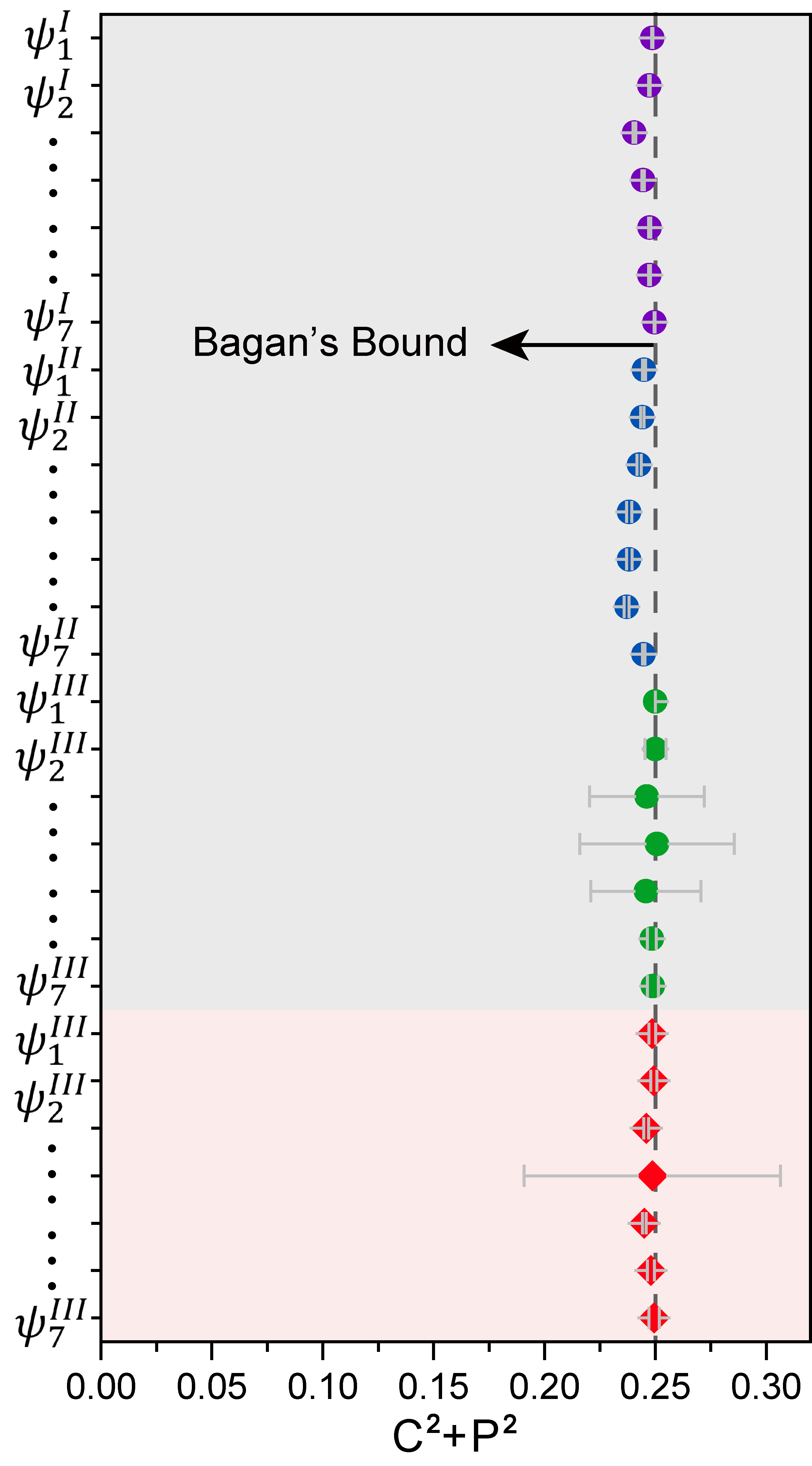}\\
 \caption{
  \textbf{The Bagan's equality defined by the duality relation between coherence and path information.} Sum of $C$ square and $P$ square of all generated states are listed and compared with the theoretical bound $0.25$, where the parts with grey (pink) background are the results obtained with state tomography (optimised positive operator valued measure (POVM)). All the states obey this new duality relation and Bagan's equality is well satisfied with acceptable errors.}
 \label{FIG. 4.}
\end{figure}

These generated states are then analysed by quantum state tomography with maximum likelihood reconstruction\cite{tomo}. From the tomographic details, we are able to calculate both $C$ and $P=P_s-\frac{1}{N}$ to estimate the wave and particle nature of the target qubit. We measure this coherence-path information duality complementarity with all three classes of states and plot the results in Fig.2. Error bars in Fig.2 are calculated with Monte Carlo simulation with the Poissonian statistics of the detection process taken into account. For all the three classes of states from Fig.2a to Fig.2c, as we tune the parameter $\zeta$, we are able to observe a continuous transition between quantum coherence and path information.

We can see that the measured $C$ and $P$ follow the theoretical curve with some small deviations. From the analytical view, the entanglement quality is reflected by the off-diagonal term of the two-qubit state density matrix. In practice, the entanglement is prepared to approach the ideal value of $0.5$. We achieve $0.49$ in our experiment. After tracing out the detector qubit, the quantum coherence quantity of the target qubit will reveal the imperfection with a small deviation. Another thing we should mention is that the entanglement quality do affect the degree of complementarity between quantum coherence and path information. From Fig.2a to Fig.2c, the three classes of states show apparent difference in degree of complementarity.

Apart from direct calculation of $P$ value from the tomographic data, we also use optimised POVM to directly measure the maximum probability of successfully identifying the state with $\{\psi^{III}_i\}$. We theoretically constructed the projective operators according to the formula in Ref.\cite{POVM}. The results are given in Fig.3a and the measured results agree with the theoretical prediction very well. Error bars in Fig.3a are directly calculated by using the propagation of error formulas. We have also compared the mismatch between the $P$ values in Fig.2c and Fig.3a, as shown in Fig.3b. Except for the result of $\psi^{III}_4$, the distances between the two different results are nearly invisible, which implies that the two methods are equivalent in principle.

Finally, we test the upper bound of this new duality relation by summing the square of both $C$ and $P$ with our experimental results. As shown in Fig.4, all the measured results meet the Bagan's equality with acceptable errors. Each $C$ and $P$ error is estimated via $1,000$ rounds of Monte Carlo simulation based on photon number detection governed by Possionian statistics. Considerable error bars come from the uncertainty when we measure rather small observables. All these experiment results faithfully prove the validation of this new duality relation between quantum coherence and path information.

\subsection*{Discussion}

We have experimentally tested the recently derived duality relation between quantum coherence and path information with polarisation encoded entangled two-qubit state. We propose and demonstrate a new way to generate different classes of partial entangled states and therefore can explore Bagan's equality for the case of $N=2$ under different conditions. Note that it is impossible to obtain the same results by using classical light in our scheme and our setup. We  become aware that Yuan {\it et al.} also test Bagan's equality with heralded single photon state\cite{Yuan2018}. A classical light field (no matter whether a weak laser or thermal light) applied on Yuan {\it et al.}'s setup will generate exactly the same results. In our experiment, we design and prepare an exotic two-photon entangled state specifically for rigorously testing Bagan's theory in the quantum regime.

Furthermore, our results show a potential connection between the two quantum resources of coherence and entanglement. Although Bagan's theory doesn't require entanglement, with different class of partial entangled quantum states, we are able to test Bagan's duality trade-off with different degree of complementary. Since the relation between coherence and entanglement remains an important and open topic\cite{StreSDB15}, our scheme and setup may serve as a new platform to simultaneously investigate the roles and their relations to coherence, path information and entanglement.

Last but not least, our scheme can be extended to larger $N$ for testing multi-path Bagan's theory and for other quantum tasks. The key operation in generating our exotic two-photon entangled states is to introduce polarisation-dependent differential transmission rates on each arm of a Bell state. Such operation can be also applied onto other multi-photon states, such as GHZ states, cluster states and even multi-photon mixed states. Especially applying our operation onto GHZ states will enable a genuine test of multi-path Bagan's theory, which will go beyond equality and test the duality between coherence and path information in an inequality fashion. There have been demonstrations of high fidelity multi-photon entanglement state\cite{ten,cluster}, thus it is experimentally feasible to generate a multi-path exotic state of our type to test Bagan's theory and to explore the relation between coherence, path information and multi-partite entanglement, even though much technical effort still remains to be made.

The wave-particle duality relations are of significance in quantum mechanics, while the quantum coherence plays a central role in quantum information processing. Our results may inspire further theoretical and experimental investigations on such fundamental researches and strengthen the prominent role of coherence in quantum physics and quantum technologies. 

\subsection*{Methods}

\textbf {Experiments.}
Entangled photon pairs are generated by using type-II spontaneous parametric down-conversion. A $5$mW $405$nm laser diode pumps a $25$mm periodically-poled KTiOPO4 (PPKTP) crystal in a Sagnac interferometer. The beam spot size at the center of the crystal is $220\mu m$. The degenerate downconversion photon pairs at $810$nm are filtered by $3$nm  bandpass filter, then coupled into single mode fibers and guided to the state preparation stage. The coincidence rate is about 30 kHz. By either inserting Brewster windows or place a PBS, three different classes of states can be generated. Photons are analysed by quantum state tomography and detected by single-photon counting modules (SPCM).

\bigskip

\subsection*{Acknowledgments}
The authors thank J.-W. Pan for helpful discussions. This work was supported by National Key R\&D Program of China (2017YFA0303700); National Natural Science Foundation of China (NSFC) (61734005, 11761141014, 11690033); Science and Technology Commission of Shanghai Municipality (STCSM) (15QA1402200, 16JC1400405, 17JC1400403); Shanghai Municipal Education Commission (SMEC)(16SG09, 2017-01-07-00-02-E00049); X.-M.J. acknowledges support from the National Young 1000 Talents Plan. S.-M. F. acknowledges support from NSFC under No. 11675113 and Key Project of Beijing Municipal Commission of Education under No. KZ201810028042. V. V thanks the Oxford Martin School, Wolfson College and the University of Oxford, the Leverhulme Trust (UK), the John Templeton Foundation, the EU Collaborative Project TherMiQ (Grant Agreement 618074), the COST Action MP1209, the EPSRC (UK) and the Ministry of Manpower (Singapore). This research is also supported by the National Research Foundation, Prime Ministers Office, Singapore, under its Competitive Research Programme (CRP Award No. NRF- CRP14-2014-02) and administered by Centre for Quantum Technologies, National University of Singapore.

\end{document}